\documentclass[twocolumn,preprintnumbers,amsmath,amssymb,superscriptaddress,prb]{revtex4}

\usepackage{graphicx}
\usepackage{dcolumn}
\usepackage{bm}
\usepackage{amsbsy}
\usepackage{amsmath}
\usepackage{amssymb}

\begin{document}
\newcommand{\sentencefrac}[2]{\displaystyle \frac{#1}{#2}}
%
\newcommand{\partderiv}[2]{\sentencefrac{\partial #1}{\partial #2}}%
\newcommand{\half}{\sentencefrac{1}{2}}
\newcommand{\Ozcan}{$\mathrm{\ddot{O}}$z\-can}
\newcommand{\via}{\emph{via}}
\newcommand{\YBCO}[1]{YBa$_{2}$Cu$_{3}$O$_{#1}$}
\newcommand{\boldYBCO}[1]{YBa$_{\mathbf{2}}$Cu$_{\mathbf{3}}$O$_{\mathbf{#1}}$}%
\newcommand{\LSCO}[2]{La$_{#1}$Sr$_{#2}$CuO$_{4}$}
\newcommand{\TiOtwo}{TiO$_{2}$}
\newcommand{\Tc}{T_{\mathrm{c}}}
\newcommand{\boldTc}{T_{\mathbf{c}}}
\newcommand{\To}{T_{0}}
\newcommand{\Bc}[1]{B_{\mathrm{c #1}}}
\newcommand{\Tesla}{\mathrm{T}}
\newcommand{\Kelvin}{\mathrm{K}}
\newcommand{\etal}{\emph{et al.}}
\newcommand{\perse}{\emph{per se}}
\newcommand{\Zs}{Z_{\mathrm{s}}}
\newcommand{\Rs}{R_{\mathrm{s}}}
\newcommand{\Xs}{X_{\mathrm{s}}}
\newcommand{\Xo}{X_{\mathrm{0}}}
\newcommand{\romanj}{\mathrm{j}}
\newcommand{\romani}{\mathrm{i}}
\newcommand{\fo}{f_{\mathrm{0}}}
\newcommand{\fB}{f_{\mathrm{B}}}
\newcommand{\fback}{f_{\textrm{back}}}
\newcommand{\muO}{\mu_{\mathrm{0}}}
\newcommand{\lambdao}{\lambda_{0}}
\newcommand{\Ao}{\mathrm{\AA}}
\renewcommand{\Re}{\textrm{Re}}
\newcommand{\complexrho}{{\rho}}
\newcommand{\rhos}{{\rho_{\mathrm{s}}}}
\newcommand{\rhom}{{\rho_{\mathrm{m}}}}
\newcommand{\rhov}{{\rho_{\mathrm{v}}}}
\newcommand{\rhoff}{\rho_{\mathrm{f\mbox{}f}}}
\newcommand{\rhon}{\rho_{\mathrm{n}}}
\newcommand{\rhoo}{\rho_{0}}
\newcommand{\omegap}{\omega_{\mathrm{p}}}
\newcommand{\Js}{J_{\mathrm{s}}}
\newcommand{\boldJs}{\boldsymbol{J}_{\mathrm{s}}}
\newcommand{\JDC}{J_{\mathrm{DC}}}
\newcommand{\Fp}{F_{\mathrm{p}}}
%
%

\title{Insulating transition in the flux-flow resistivity\\
of a high temperature superconductor}

\author{Benjamin Morgan}
 \email{B.Morgan.96@cantab.net}
\affiliation{Department of Physics, Cavendish Laboratory, J J Thomson Avenue,
             Cambridge, CB3 0HE, United Kingdom}
%
\author{D.~M.~Broun}
\affiliation{Physics Department, Simon Fraser University, Burnaby, BC, V5A~1S6, Canada}
\author{Ruixing Liang}
\author{D.~A.~Bonn}
\author{W.~N.~Hardy}
\affiliation{Department of Physics and Astronomy, University of British Columbia, Vancouver,
             BC, V6T~1Z1, Canada}
\author{J.~R.~Waldram}%
\affiliation{Department of Physics, Cavendish Laboratory, J J Thomson Avenue,
             Cambridge, CB3 0HE, United Kingdom}
\date{\today}

\begin{abstract}
Measurements of the DC resistivity of under-doped cuprate
superconductors\cite{Ando_logarithmic_LSCO_resistivity,Boebinger_logarithmic_LSCO_resistivity}
have revealed a metal--insulator transition at low temperatures when superconductivity is
suppressed by a very large magnetic field, with the resistivity growing logarithmically in the
low temperature limit.
This insulating behaviour has been associated not only with the large magnetic fields,
but also with the under-doped composition and intrinsic sample inhomogeneity,
and it is important to establish whether these factors are essential to it.
Here we report high resolution microwave measurements of the flux-flow resistivity
of optimally doped \YBCO{6+x} in the mixed state at temperatures down to 1.2~K.
We find that the effective resistivity of the vortex cores exhibits a
metal--insulator transition, with a minimum at 13~K and a
logarithmically growing form below 5~K.
The transition is seen in samples of the highest quality and in magnetic fields as low as 1~T.
Our work is the first report of a metal--insulator transition in
optimally doped \YBCO{6 + x}, and the first such transition to be seen in a system in which
superconductivity has not been globally suppressed.
\end{abstract}

\maketitle

\begin{figure*}
\includegraphics*[width=\textwidth]{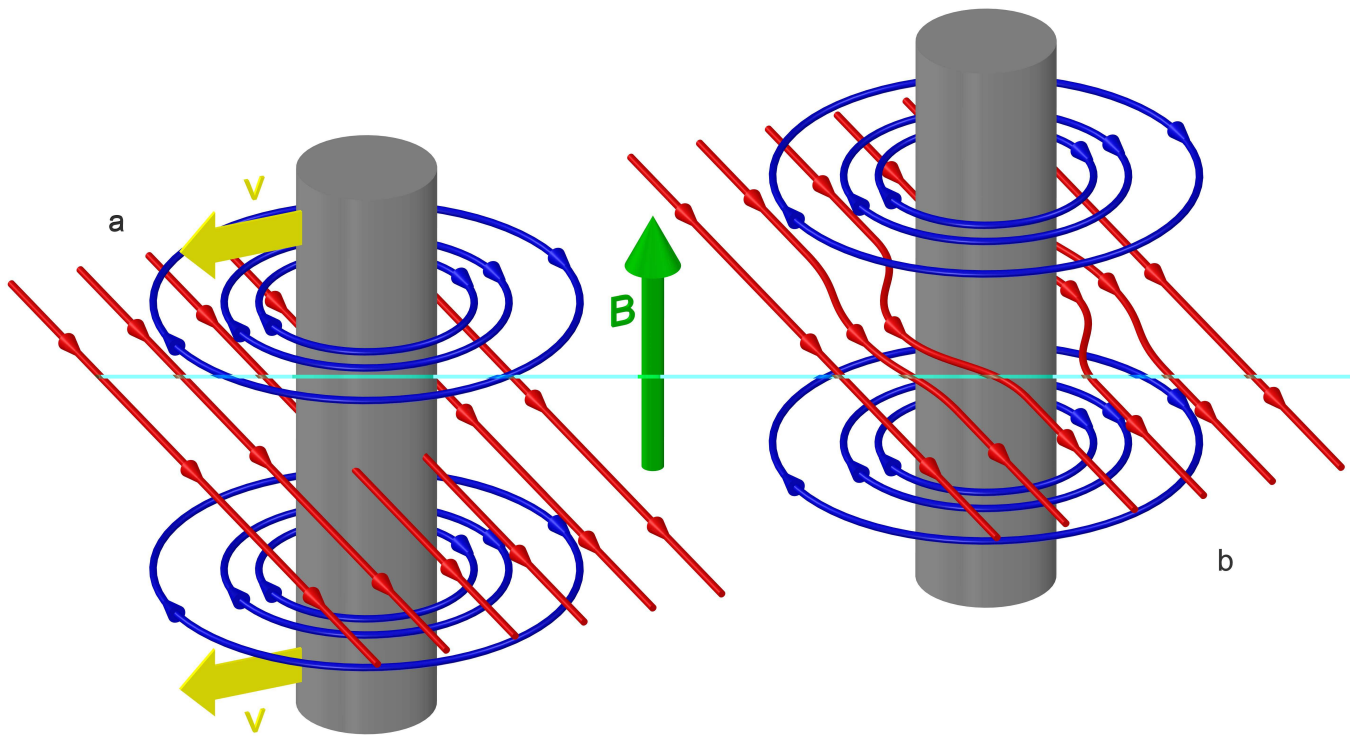}
\caption{A vortex in a superconductor in an applied DC magnetic field $B$ (green arrow),
responding to a DC electric current.
The core of the vortex is represented by a cylinder of normal metal through which a
line of magnetic flux passes. In order to screen this flux from the
bulk of the superconductor, supercurrents of density $\Js$ flow around the normal core
(blue arrows).
The external application of a DC current density $\JDC$ (red arrows)
causes a Lorentz force to be exerted upon the vortex.
In (a) the vortex is free to respond to the force and moves perpendicularly to the applied
current and the magnetic field with a velocity $v$.
This movement of the vortex generates an electric field in the core region \via{}
the Faraday effect and the applied current flows through the core,
causing electrical dissipation.
In (b) the vortex is pinned in place by a material inhomogeneity.
Since the vortex does not move, no electric field exists to drive the applied
current through the core. In this case, the surrounding superconductor acts as a
``short circuit'' through which all of the DC electric current flows,
resulting in no dissipation and no measurable flux-flow resistivity.}
\label{figure1}
\end{figure*}
Any theoretical description of a superconductor must be rooted in an understanding of the
ground state from which superconductivity arises, information about which can be obtained by
measuring the electrical resistivity in the normal state of the material.
However, the high magnetic fields required to suppress superconductivity in
the cuprate superconductors make it extremely difficult to measure
the normal state resistivity in the low temperature limit.
The most widely studied cuprate superconductor is optimally doped
\YBCO{6 + x} but its upper critical magnetic field\cite{Sekitani_YBCO_Hc2}
$\Bc{2}$ is in excess of 100~T, which is extremely difficult to generate and
impossible to maintain in the laboratory.
However, superconductivity can be suppressed
in localized regions\cite{Tinkham_superconductivity}
of a sample by applying a field greater than its
lower critical field $\Bc{1}$,
which is\cite{Liang_YBCO_Hc1} of order 0.1~T for optimally doped \YBCO{6 + x}.
This gives rise to the mixed state,
in which the superconducting sample is threaded by lines of magnetic flux
surrounded by vortices of supercurrent.
In the vicinity of each vortex, electrons are removed from the
superconducting condensate and become subject to normal resistive processes:
they may be thought of as a quasi-normal metallic system.

\begin{figure}
\includegraphics{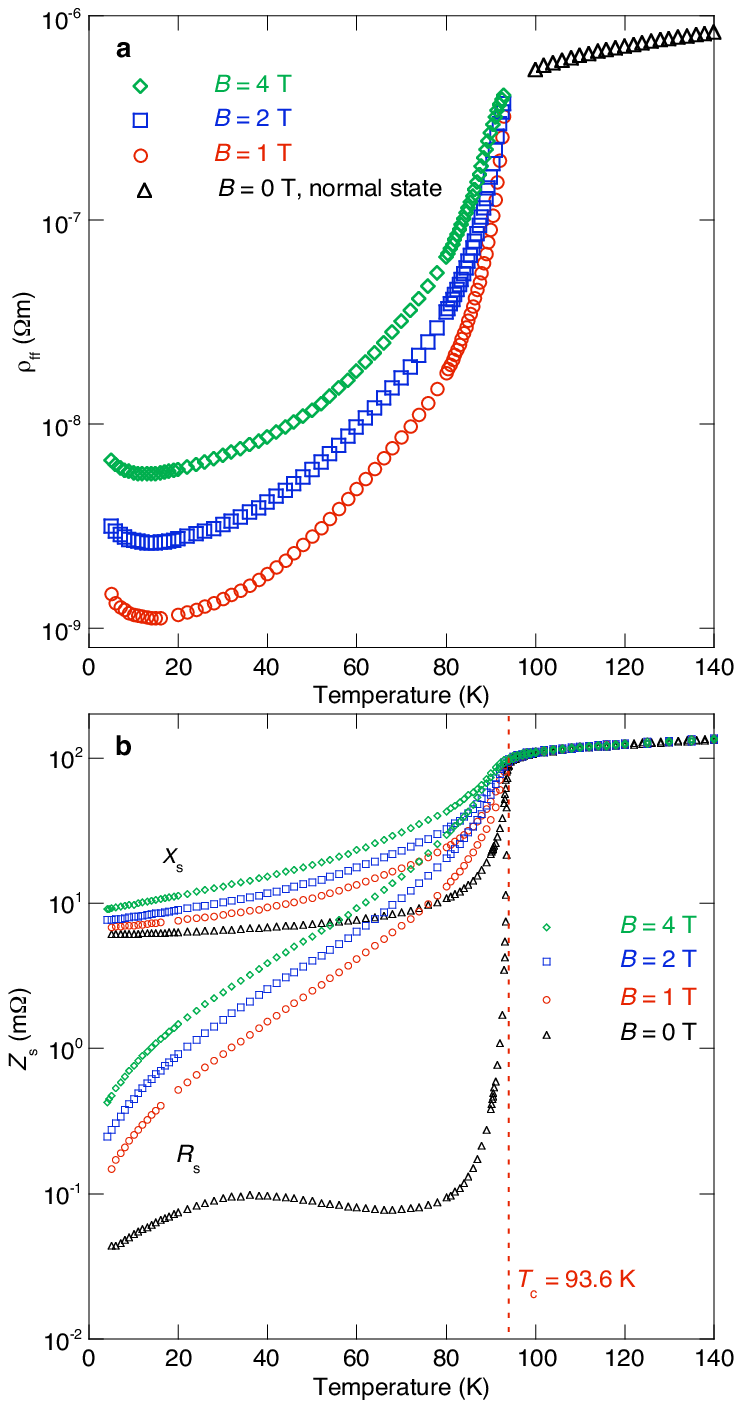}
\caption{Microwave measurements of optimally doped
\YBCO{6 + x} in DC magnetic fields of 1~T (circles), 2~T (squares)
and 4~T (diamonds) and in zero field (triangles).
(a) The flux-flow resistivity $\rhoff(T)$ plotted on a logarithmic scale.
Also shown is the real part of the complex resistivity in the normal state (triangles),
to which $\rhoff$ tends as $T \to \Tc$.
In all three data sets, $\rhoff(T)$ has a minimum at $T = 13$~K.
The magnetic field dependence of $\rhoff$ at low temperatures is predominantly due to the
increase in the proportion of the sample that is occupied by the vortices
as the magnetic field is increased.
(b) The surface impedance data from which the flux-flow resistivity data were extracted.}
\label{figure2}
\end{figure}
If a transport current is passed through the sample
and the vortices are free to move in response to it,
the induced electric field will drive the transport current through the
vortex core regions\cite{Bardeen_Stephen_flux_flow}, as illustrated in Figure~1(a).
The current will be carried by non-superconducting electrons and subjected to resistive
dissipation.
This process can be described by a measurable quantity,
the flux-flow resistivity $\rhoff$, which is the effective resistivity
of the vortex cores multiplied by the fraction of the material
that they occupy. Figure~2(a)
shows microwave measurements of $\rhoff(T)$ on a high-quality sample of
optimally doped \YBCO{6 + x}.
At a temperature close to the superconducting transition temperature $\Tc$,
the flux-flow resistivity joins smoothly with the normal state resistivity.
Just below $\Tc$ it falls sharply,
because $\rhoff$ is weighted according to the size of the vortex cores,
which shrink dramatically as the temperature is reduced just below the superconducting
transition. At low temperatures, the flux-flow resistivity exhibits
a minimum in strength at $T = 13 $~K and a pronounced increase as
the temperature is lowered further.
At these temperatures, the core size is almost independent of
temperature\cite{mSR_vortex}, so the observed form of $\rhoff(T)$ must reflect
the temperature dependence of the core resistivity itself.

At first sight it is surprising that the low temperature upturn in $\rhoff(T)$ has not been
observed before, since optimally doped \YBCO{6 + x} is the most studied
high-$\Tc$ superconductor.
However, although in a perfect sample the vortices would be free to move in response to
an applied DC transport current, in real samples at low temperatures
the vortices are strongly pinned by local material inhomogeneities at which
superconductivity is suppressed.
In low frequency experiments the flux can not flow, there is no induced electric field and
no current is driven through the vortex core.
Instead, the current flows around the core region, as illustrated in Figure~1(b),
with the result that there is no dissipation and the flux-flow resistivity can not be measured.
Nevertheless it \emph{can} be measured in a high frequency experiment,
even in the presence of vortex pinning.
In response to a microwave transport current,
the vortices oscillate about their pinned positions at the microwave frequency $\omega$,
an electric field is induced and the current flows through the normal cores.
However, the effective resistivity associated with vortex motion is no longer simply
the flux-flow resistivity $\rhoff$,
but a complex quantity $\rhov$ which is determined by the effects of both
resistive flux flow and vortex pinning\cite{Gittleman_Rosenblum_JApplPhys}.
It is given by the formula\cite{Tinkham_superconductivity}
$\rhov = \rhoff \left( 1 - \romani \frac{\omegap}{\omega} \right)^{-1}$
where $\omegap$ is known as the de-pinning frequency.

Several research groups\cite{Morgan_Flux_flow, YBCO_thin_film_depinning_frequency,
Parks_YBCO_mixed_THz, YBCO_vortices_electronic_state_Zs} have demonstrated the effectiveness
of microwave techniques in measuring the flux-flow resistivity of high-$\Tc$ superconductors
at temperatures as low as 10~K.
At such low temperatures, the complex resistivity of the sample as a whole
is well described by\cite{Coffey_Clem_PhysRevLett,Brandt_TypeII_AC}
$\rhom = \rhov + \rhos$ where $\rhos$ is the complex resistivity
of the superconducting regions in between the vortices, which can reasonably be assumed to be
the same as that measured with the sample in zero DC magnetic field, since $B \ll \Bc{2}$.
Thus to extract $\rhov$ (and hence $\rhoff$) from a microwave measurement,
the \emph{full complex resistivity} of the sample measured in zero field
must be subtracted from that measured in the mixed state.
Until recently however, it has not been feasible to
resolve the detailed temperature dependence of the flux-flow resistivity,
especially at very low temperatures.
For the experiments reported here, a new technique was developed\cite{My_thesis}
that allows these measurements to be made with unprecedented precision\cite{Huttema_RSI}.
The success of the innovation is apparent in the surface impedance data of Figure~2(b),
which show negligible random noise.

In order to compare the results of the present work with
the DC resistivity measurements in magnetic fields of up to 60~T by Ando and
Boebinger\cite{Ando_logarithmic_LSCO_resistivity,Boebinger_logarithmic_LSCO_resistivity},
it will be useful to convert the quantity
$\rhoff(T)$ into an effective normal resistivity $\rhon(T)$ associated with the
non-superconducting electrons in the vortex core regions.
A standard method to achieve this\cite{Strnad_Flux_flow_PRL_1964}
is to use the empirical relation
$\frac{\rhoff}{\rhon} \sim \frac{B}{\Bc{2}}$, which is
known to describe accurately the temperature dependence of the flux-flow resistivity
of conventional superconductors in the limits
$\frac{T}{\Tc} \to 0$ and $\frac{B}{\Bc{2}} \to 0$.
Note that the function we use for $\frac{B}{\Bc{2}(T)}$ shows a very weak
temperature dependence at low temperatures\cite{Sekitani_YBCO_Hc2}, which is
consistent with published measurements of the vortex core size\cite{mSR_vortex}.
The temperature dependence of $\rhon(T)$ can also be seen in the $\rhoff(T)$ data,
which are effectively only re-scaled at low temperatures by an almost constant factor.
Figure~3 shows the resulting $\rhon(T)$ from an experiment in which the temperature
range was extended down to 1.2~K.
In Figure~3(b) the data are plotted on a logarithmic temperature scale.
At temperatures below 5~K the quasi-normal resistivity satisfies the relation
$\rhon = \rhoo \ln \left( \frac{\To}{T} \right)$ where $\rhoo$ and $\To$ are constants.
The same logarithmic expression describes the low-temperature resistivity of 
under-doped \LSCO{2-x}{x} in which superconductivity has been globally suppressed
by a very large magnetic field\cite{Ando_logarithmic_LSCO_resistivity,
Boebinger_logarithmic_LSCO_resistivity}.

\begin{figure}
\includegraphics{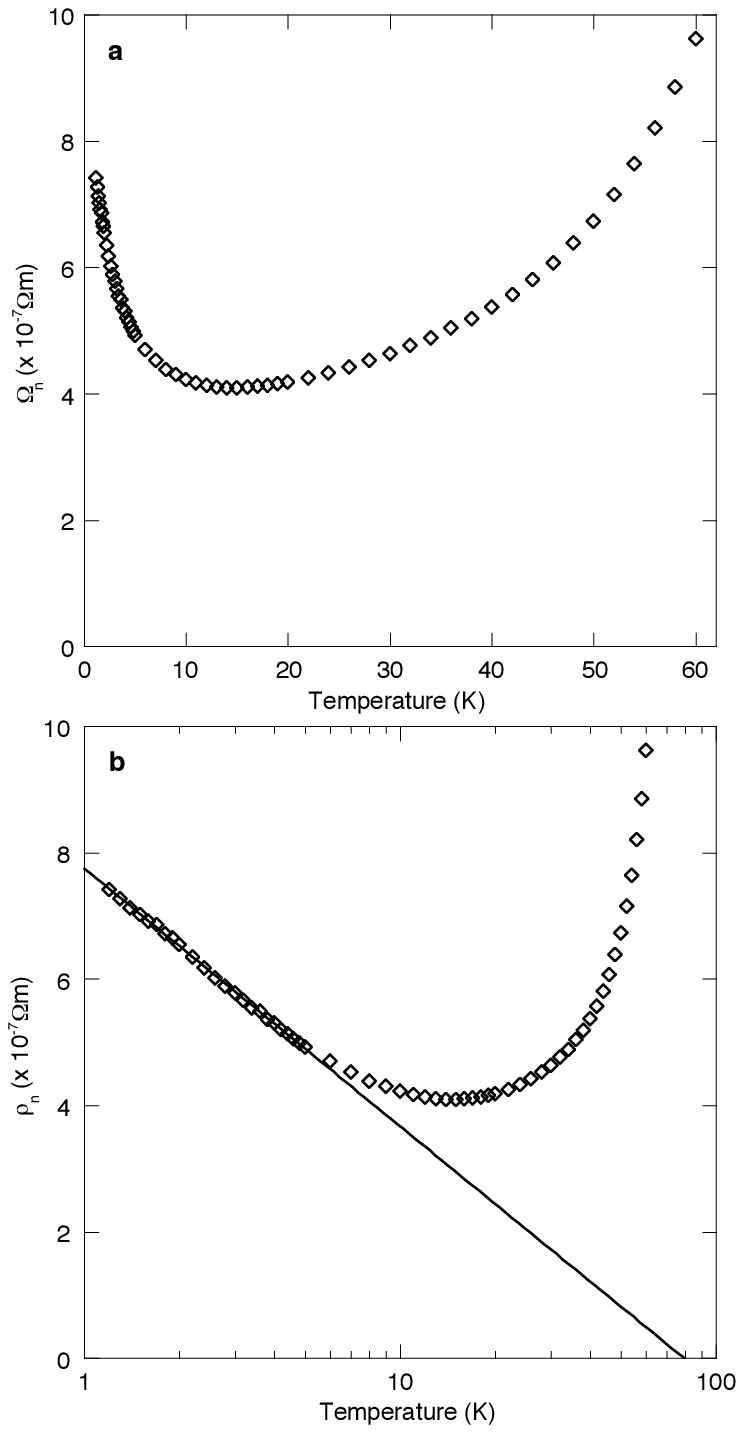}
\caption{The effective resistivity $\rhon(T)$ of the normal cores of vortices in
optimally doped \YBCO{6 + x} in a magnetic field of 4~T,
found using $\rhon(T) = \rhoff(T) \frac{B}{\Bc{2}(T)}$
and taking $\Bc{2}(T)$ from a smooth fit to the data of Reference \cite{Sekitani_YBCO_Hc2}.
Note that this empirical relation\cite{Strnad_Flux_flow_PRL_1964}
can not set the absolute magnitude of $\rhon$,
whose scale is uncertain by an arbitrary constant factor.
Only the lowest temperatures are shown in order to examine the upturn below 13~K
in more detail.
(a) Plotted on a linear temperature scale.
(b) Plotted on a logarithmic temperature scale,
clearly showing that $\rhon \propto \ln \left( \frac{1}{T} \right)$
for $T \leq 5$~K.}
\label{figure3}
\end{figure}
These results make several new and important contributions
to our understanding of high-$\Tc$ superconductors.
The metal--insulator transition observed in our work occurs in
ultra-high purity samples, at low magnetic fields and at optimal doping.
These features place tight constraints upon theoretical explanations of the
metal--insulator transition in the cuprates: it can no longer be viewed as a consequence
simply of disorder or high fields, or as confined to the under-doped part of the phase diagram.
We have shown that the insulating response with
$\rhon \propto \ln \left( \frac{1}{T} \right)$ persists at least up to optimal doping
in the \YBCO{6 + x} system, a doping level at which there is
little evidence\cite{Orenstein_review} of magnetic order, charge order or
granular metallic behaviour.
Moreover, possibly the most significant new feature of the observed insulating behaviour is
that it coexists with superconductivity in the bulk material.
Low temperature measurements of the DC resistivity of cuprates necessarily rely on the global
suppression of superconductivity, either by a large magnetic
field\cite{Ando_logarithmic_LSCO_resistivity,Boebinger_logarithmic_LSCO_resistivity}
or by chemical doping\cite{Sun_Insulating_YBCO}.
Until now there existed the possibility that
superconductivity was being destroyed \emph{coincidentally} with a
metal--insulator transition in the electronic ground state,
induced by disorder or a large magnetic field.
The present work shows unambiguously that the intrinsic ground state
underlying superconductivity in optimally doped \YBCO{6 + x}
is an electrical insulator at low temperatures.
That superconductivity can arise from an insulating ground state confirms the
extremely unconventional nature of high-$\Tc$ superconductivity.

\bibliographystyle{naturemag}
\bibliography{Ordered}

\end{document}